# High Performance $TiO_2$ Ferroelectric Field Effect Transistors with $HfZrO_2$ for Neuromorphic Computing

Chandan Samanta*, Elia Palmese*, Ziyu Ouyang, Tuofu Zhama, Robinson Pino, Yuping Zeng

***Abstract*— $TiO_2$ ferroelectric field effect transistors (FeFETs) with $HfZrO_2$ (HZO) ferroelectric dielectric layers and bottom gate topology are fabricated for applications in neuromorphic systems. Two sets of devices are fabricated with different gate topologies by varying the thickness of the ferroelectric gate stack. Different device architectures are studied by varying the source drain length ($L_{SD}$) and gate length ($L_G$). The devices have high on/off ratios up to $10^7$ with low leakage off currents $<10^{12}$ A. Repeated cycle testing shows high reliability and a stable memory window. The devices have large memory windows ranging from 3 to 8 V.**

## Introduction

Traditional computers are based on the von Neumann model, where memory and processing are separated, causing energy- and time-intensive data transfer (the so-called "von Neumann bottleneck"). In contrast, neuromorphic systems are inspired by the human brain, where memory and computation are co-located in networks of neurons and synapses. This brain-like approach enables massive parallelism, adaptability, and ultra-low power operation. Neuromorphic computing holds great promise for tasks that require real-time learning, pattern recognition, and sensory data processing—such as autonomous driving, robotics, edge computing, and AI.

Current existing neuromorphic computing systems (Intel’s Loihi, IBM’s TrueNorth, Heidelberg University’s BrainScale S) use conventional CMOS technology. While CMOS has enabled successful neuromorphic chips like Intel’s Loihi and IBM’s TrueNorth, it faces fundamental disadvantages for brain-inspired computing. Synaptic weights must be stored in SRAM or DRAM and moved to logic for computation, creating a von Neumann bottleneck that wastes energy and limits speed. Emulating a single synapse requires many transistors, leading to large area, high power, and poor scalability compared to the brain’s ultra-dense networks. Moreover, CMOS is inherently digital, making analog operations such as in-memory matrix–vector multiplication or spike-timing dependent plasticity inefficient.

---

We thank the University of Delaware Nanofabrication Facility staff for their assistance and support. The work is supported by the Department of Energy with No. DE-SC0025379, DE-SC0026053, DE-SC0026335, and partially supported by National Science Foundation with No. 2239302 and National Aero and Space Administration with No. 80NSSC22M0171(Corresponding author: Yuping Zeng, yzeng@udel.edu).

C. Samanta, E. Palmese, Ouyang Ziyu, Tuofu Zhama, Yuping Zeng are from the Department of Electrical and Computer Engineering at University of Delaware. Robinson Pino is affiliated with United Semiconductors.

*C. Samanta and E. Palmese equally contribute to this work.

Ferroelectric field-effect transistors (FeFETs) have emerged as a promising platform for neuromorphic computing by integrating nonvolatile memory functionality with energy-efficient logic within a simple transistor structure [1-3]. Although the practical deployment of FeFETs is still in its early stages, extensive research has demonstrated their potential using silicon [4, 5], $MoS_2$ [6-8], III–V semiconductors [9, 10], and IGZO [11, 12]. In contrast, $TiO_2$-based FeFETs have been much less explored. Ref. 13-15 [13-15] reported $TiO_2$ FeFET results, but employed $HfO_2$ or HfLaO as the ferroelectric, achieving only small memory windows of 1.1–2.4 V and on/off ratios below $10^5$. Other reports of $HfZrO_2$ (HZO) $TiO_2$ FeFETs also show small memory windows and low on/off ratios below $10^5$ [16].

Our group has previously demonstrated top-gated $TiO_2$ thin-film transistors with record-high performance [17-20]. Here, we extend this work by developing $TiO_2$ FeFETs in a bottom-gate configuration, incorporating HZO as the ferroelectric dielectric to realize metal–ferroelectric–metal–insulator–semiconductor (MFMIS) like FeFET structures.

## Experimental Methods

Two sets of samples are fabricated for this paper, set A and set B. The device fabrication starts by solvent cleaning Si substrates with 260 nm of $SiO_2$ using NMP, acetone, and isopropanol. A 30 nm of titanium nitride (TiN) layer is blanket deposited via sputtering. The TiN layer is patterned and dry and wet etched to form the bottom gate electrode. Dry etching is done in a fluoric inductively couple plasma (F-ICP) etcher with $CF_4$/Ar chemistry. The samples are dipped in a $NH_4OH$, $H_2O_2$, and $H_2O$ solution (1:4:20) heated at 40 °C for 20 seconds to remove any remaining TiN residue. Next, a HZO layer is deposited on the samples at 250 °C via thermal atomic layer deposition (ALD). The HZO thickness for set A is 10 nm and 8 nm for set B. Devices in set A are subjected to wet etching in 5 % hydrofluoric acid (HF) for 40 seconds to remove the HZO layer from the TiN gate region. The HZO layer on the TiN gate is left unetched for set B. Next, the samples are annealed in $N_2$ for 1 minute at 350 °C. After annealing, 8 nm of $HfO_2$ is deposited on both samples at 250 °C followed by 15 nm of $TiO_2$ deposited at 150 °C via thermal ALD. To improve the mobility of the $TiO_2$ the samples are annealed for 30 minutes in $O_2$ at 375 °C for set A and 500 °C for set B. Next, $TiO_2$ mesa etching is completed using a Fl-ICP. After mesa etching, the $HfO_2$ layer is removed from the TiN gate using 5 % HF for 40 seconds for set A and again left unetched for set B. Ohmic source and drain contacts are formed with 250 nm of Aluminum. Finally, set B has a 15/150 nm layer of Ni/Au deposited on the TiN gates with varying overlap areas. Figure 1 illustrates the fabrication process flow for the FeFETs with the device architecture at different steps including process variation between sets A and B in blue.

A Keysight B1500 semiconductor parameter analyzer is used to measure transfer curves, ($I_D$–$V_{GS}$), and capacitance versus voltage (CV) measurements of the fabricated FeFETs and test structures.

## Results and Discussion

To characterize the ferroelectric properties of our HZO layers, simple metal-ferroelectric-metal (MFM) capacitors are fabricated. These devices have a TiN bottom contacts followed by 10 nm of HZO annealed at 350 °C and circular Ni/Au top contacts. To characterize the ferroelectric property of the MFM capacitors CV measurements are performed. Figure 2 shows the CV measurement taken on a MFM capacitor from -5 to 5 V with a frequency of 5 kHz. In the inset is a cross section of the MFM capacitor. The CV curve shows a butterfly shape commonly associated with capacitance measurements for ferroelectric materials due to polarization changes at different applied voltages.

The first set of FeFETs, set A, were fabricated with the intention of entirely removing the HZO and $HfO_2$ oxide from the TiN gate. After fabrication, electrical testing revealed that a thin insulating oxide layer remains on the TiN gate pad which enhances the ferroelectric memory window of the FeFETs. To investigate the mechanism behind the ferroelectric window enhancement, a second set of FeFETs, set B, are fabricated intentionally leaving the HZO and $HfO_2$ on the TiN gate to determine how the oxide thickness on the TiN gate influences the memory window. Additionally, Ni/Au top contacts are deposited on set B devices to form a MFMIS like structure and study the influence of overlap area between top Ni/Au and the bottom TiN on the ferroelectric properties of the transistors.

To characterize the behavior of the $TiO_2$ FeFETs from set A, $I_D$–$V_{GS}$ curves are measured to determine the on/off ratio and memory window of the devices. The role of source–drain distance ($L_{SD}$) and gate length ($L_G$) is studied. Figure 3 shows the dual sweep $I_D$–$V_{GS}$ measured from -7 V to 5 V with $V_{DS}$=1 V for (a) a non-overlapping gate device with $L_{SD}$=6 μm and $L_G$=3 μm and (b) an overlapping gate device with $L_{SD}$=3 μm and $L_G$=6 μm. Both devices show counterclockwise hysteresis expected for ferroelectric transistors. The memory windows for all plots are measured at $I_D$=1x10$^{-8}$ A. The devices with non-overlapping gate structure have a memory window of 4.73 V and on/off ratio greater than 2.67x10$^6$. The overlapping gate devices have a memory window of 3.3 V and on/off ratio greater than 2.27x10$^7$. The off current for both devices is taken to be 1x10$^{-12}$ A but is limited by the noise floor of the semiconductor parameter analyzer. The non-overlap devices show a larger memory window likely related to capacitive effects. Reducing $L_{SD}$ from 6 μm to 3 μm increases the on-current due to the reduced channel resistance, resulting in an on/off ratio greater than 10$^7$, compared to 10$^6$ for longer channels.

To test the stability of the ferroelectric window, cycle testing is completed on the FeFETs for both architectures. Endurance measurements show that gate overlapped devices ($L_{SD}$=3 μm/$L_G$=6 μm) exhibit superior stability under repeated cycling compared to non-overlapped devices ($L_{SD}$=6 μm/$L_G$=3 μm), with significant on-current degradation. Figure 4 (a) shows 1000 consecutive cycles of the dual sweep $I_D$–$V_{GS}$ for the non-overlap devices between -5 and 5 V with $V_{DS}$=1 V. Figure 4 (b) shows 1300 consecutive cycles the dual sweep $I_D$–$V_{GS}$ for the overlap devices between -7 and 5 V with $V_{DS}$=1 V. Figure 4 (c) and (d) track the on- and off- current as a function of cycle number for the non-overlap and overlap devices respectively. In the first 100 cycles of the non-overlap device measurements there is a multiple order of magnitude drop in on current that stabilizes around 1x10$^{-9}$ A for the remaining cycles. Testing for the overlap devices confirms endurance over 1300 cycles, with stable memory window, on-current, and low off-current (~10$^{-12}$ A), indicating low standby power.

Next, non-overlap devices from set B are measured with $L_{SD}$ of 6 μm and $L_G$ of 3 μm. Ni/Au top metal are deposited with different overlap areas on the TiN ranging between 100 μm$^2$ and 500 μm$^2$. Figure 5 (a) plots the dual sweep $I_D$–$V_{GS}$ from -12 to 6 V with $V_D$ = 1 V for an FeFET with a Ni/Au top contact that has an overlap area of 10 x 15 μm. The device has an on/off ratio of 2.96x10$^6$ ($I_{off}$ = 1x10$^{-12}$ A) with a memory window of 8.12 V. The forward and reverse subthreshold slopes are calculated to be 290 mV/dec and 183 mV/dec respectively. Preliminary cycle testing for the non-overlap devices in set B does not show a reduction in on current observed in the devices in set A. This indicates keeping the oxide layers on the gate increases device stability over time.

Utilizing a thicker ferroelectric layer on the TiN gate almost doubles the ferroelectric memory window from 4.73 V to 8.12 V. Additionally, changing the overlap area of the Ni/Au top contact affects the ferroelectric window of the devices. Figure 5 (b) plots the dual sweep $I_D$–$V_{GS}$ from -12 to 6 V with $V_D$ = 1 V for an FeFET with six different overlap gate areas ranging from 100 μm$^2$ and 500 μm$^2$. Increasing the overlap area of the top contact decreases the memory window. Initially, increases in the overlap area up to 225 μm$^2$ only shifts the turn-on threshold voltage to more negative values. Increasing the overlap area above 225 μm$^2$ results in both the on and off threshold voltages pinching inward toward a center point almost halfway between the initial memory window at ~ -3 V. The decrease in ferroelectric window is attributed to capacitance mismatch with the MFM capacitor ($C_{FE}$) and MIS gate capacitance ($C_{MIS}$). Increasing the area of the top contact increase $C_{FE}$ resulting in a larger capacitance than the $C_{MIS}$. This capacitance mismatch decreases the polarization window and is an important parameter for stable memory windows [21, 22].

## Conclusion

$TiO_2$ FeFETs using HZO as the ferroelectric layer with high on/off currents and large stable memory windows are achieved. The effects of device architecture and HZO thickness on the FeFET performance are determined.

## Figures

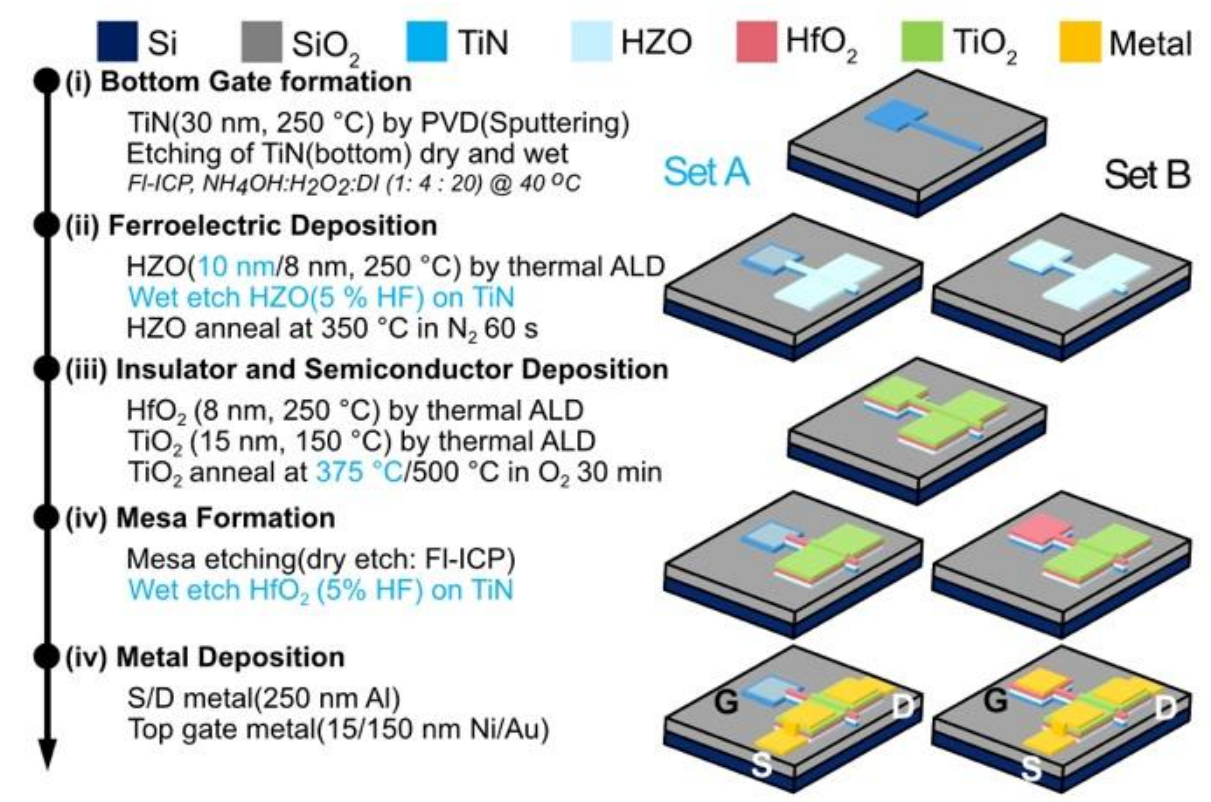

Fig. 1 Fabrication flow for bottom gated $TiO_2$ FeFETs for devices with (set A) and without (set B) gate oxide etching on the TiN gate.

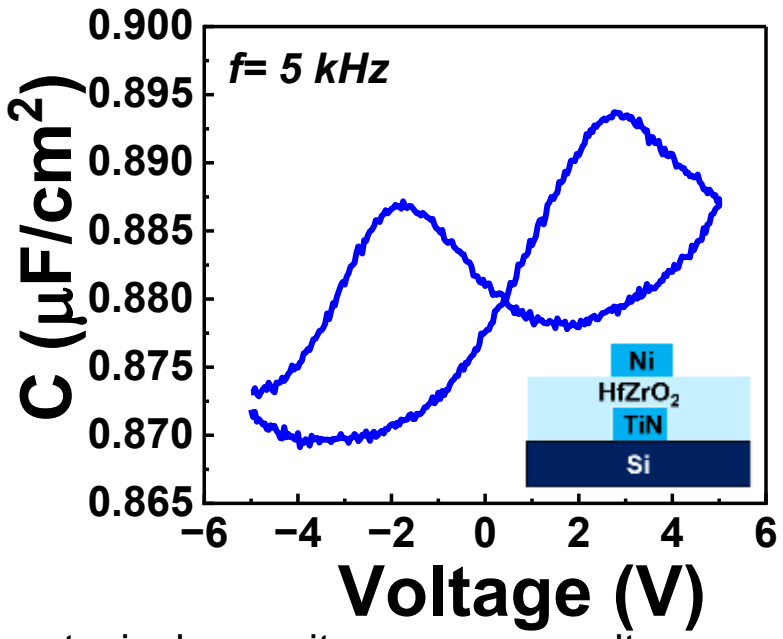

Fig. 2 Shows a typical capacitance versus voltage measurement for an MFM capacitor measured from -5 to 5 V. The inset shows a cross section of the device with TiN as a bottom contact, HZO as the ferroelectric layer, and Ni as a top contact.

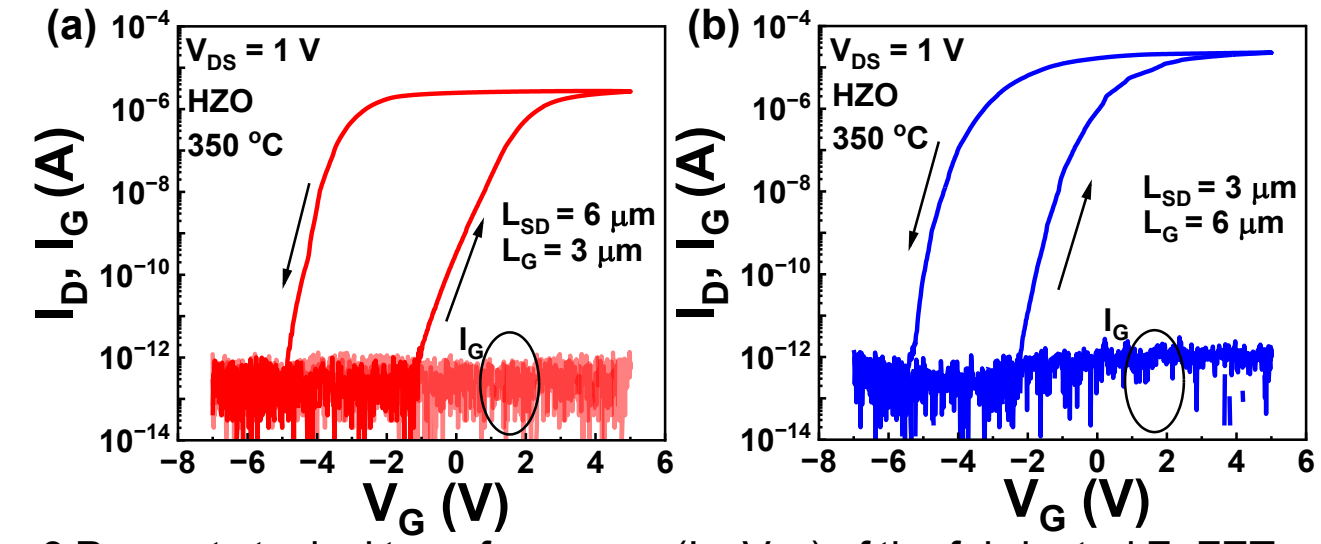

Fig. 3 Presents typical transfer curves ($I_D$–$V_{GS}$) of the fabricated FeFETs annealed at 350 °C for two different device topologies. (a) Shows the transfer curve for a non-overlap gate topology with $L_{SD}$ = 6 μm and $L_G$ = 3 μm while (b) shows a device with overlap gate topology with $L_{SD}$ = 3 μm and $L_G$ = 6 μm. Devices exhibit on/off ratios in the range of $10^6$ ~$10^7$ and memory windows of 4.73 and 3.3V respectively.

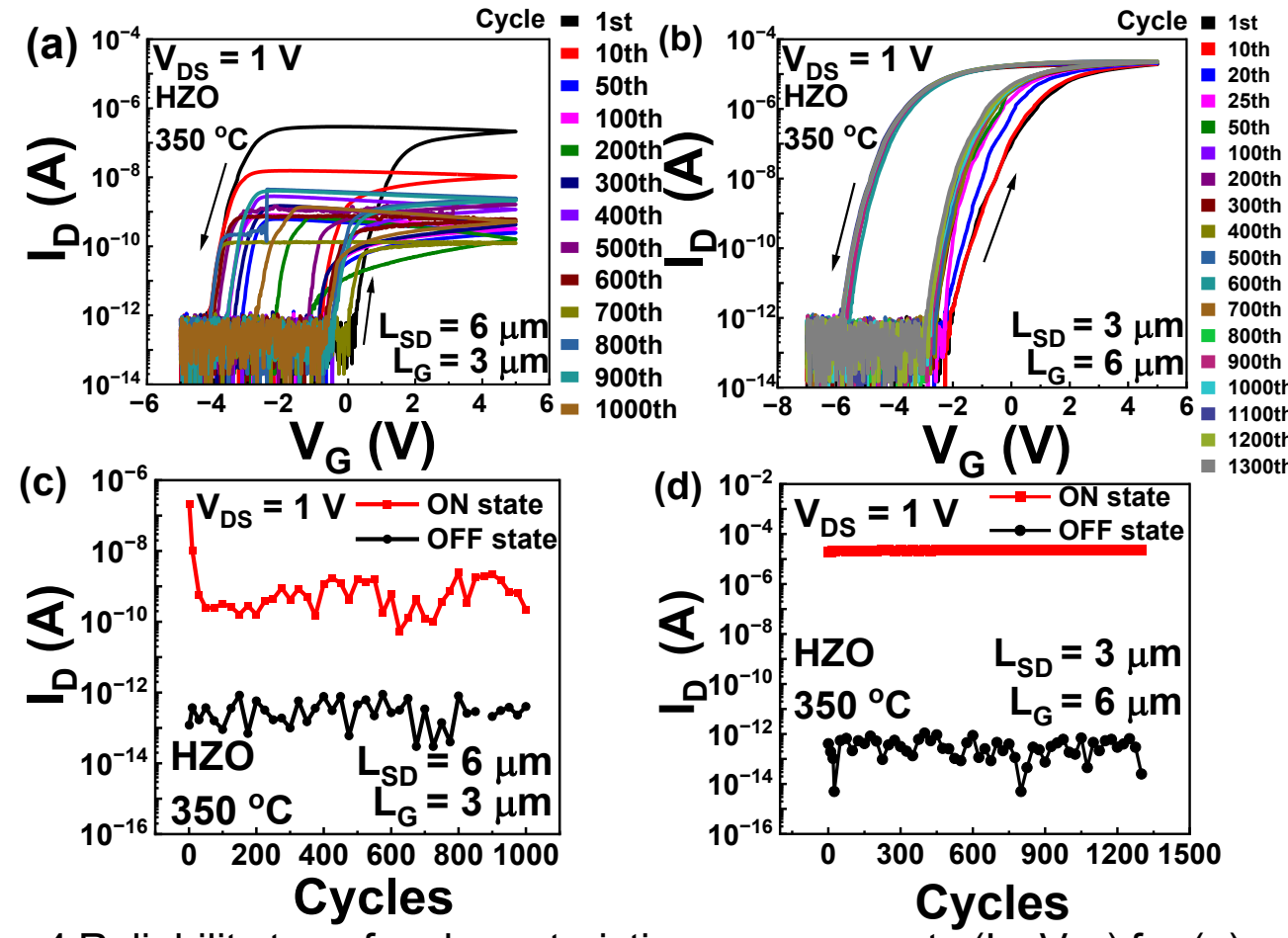

Fig. 4 Reliability transfer characteristic measurements ($I_D$–$V_{GS}$) for (a) a non-overlap device measured from -5 to 5 V for 1000 cycles and (b) an overlap device measured from -7 to 5 V for 1300 cycles. (c)-(d) Track the on and off state current at incremental points throughout the cycle testing for the non-overlap and overlap devices respectively.

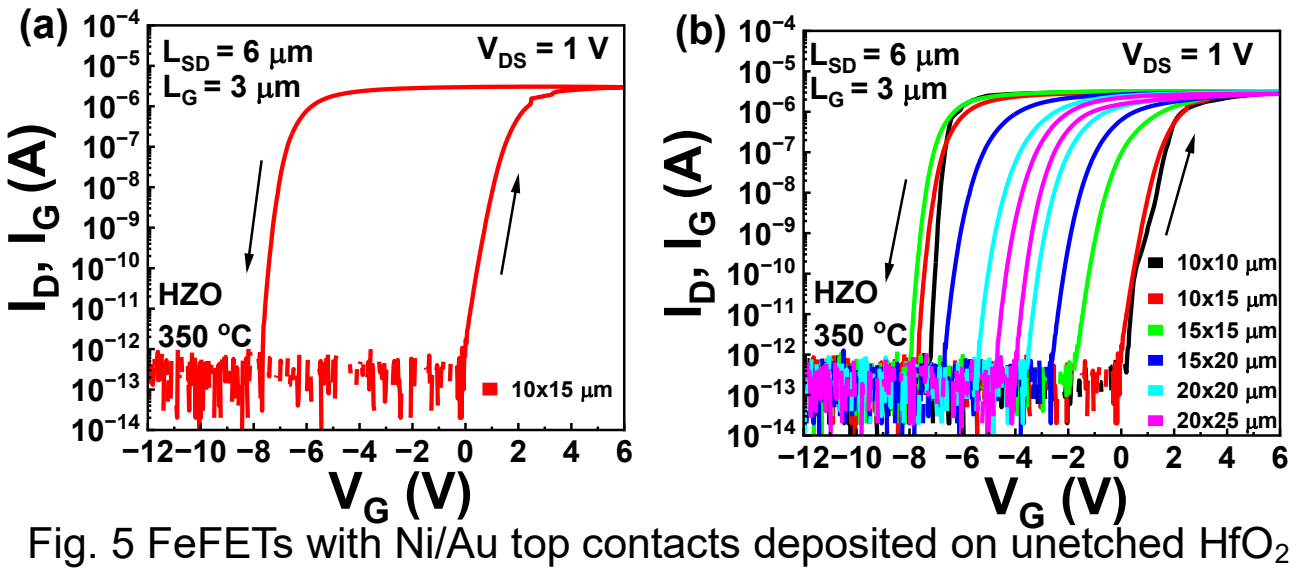

Fig. 5 FeFETs with Ni/Au top contacts deposited on unetched $HfO_2$ and HZO layers with TiN bottom gates. (a) Plots the transfer characteristics ($I_D$–$V_{GS}$) measured -12 to 6 V for a device with a Ni/Au overlap area of 10 x 15 µm. (b) Plots $I_D$–$V_{GS}$ for the FeFET with increasing overlap area up to 20 x 25 µm.